\documentclass[12pt]{article}
\usepackage{amsfonts}
\usepackage{amsmath,epsfig}
\usepackage{color}
\def\halft{{\textstyle\frac{1}{2}}}
\def\quart{{\textstyle\frac{1}{4}}}

\begin{document}
\numberwithin{equation}{section}

\title{Inhomogeneous ``longitudinal'' circularly-polarized plane waves\\
        in anisotropic elastic crystals}
\author{Philippe Boulanger,
         Michel Destrade, 
         Michael A. Hayes}
\date{}
\maketitle

%
\begin{abstract}

Conditions on the elastic stiffnesses of anisotropic crystals are
derived such that circularly polarized longitudinal
inhomogeneous plane waves with an iso\-tro\-pic slowness
bivector may propagate for any given
direction of the normal to the sagittal plane.
Once this direction is chosen, then the wave speed, the direction of
propagation, and the direction of attenuation are expressed in terms
of the mass density, the elastic stiffnesses, and the angle between
the normal to the sagittal plane and the normals
(also called ``optic axes'')
to the planes of central circular section of a certain ellipsoid.
In the special case where this angle is zero, and in this special case
only, such waves cannot propagate.

\end{abstract}

\section{Introduction}

In classical linearized elasticity theory, a special role is played by
infinitesimal longitudinal homogeneous plane waves. 
For such waves the amplitude vector
is parallel to the propagation direction $\mathbf{n}$ (say) so that all
the particles oscillate along that direction $\mathbf{n}$ and
the motion is one-dimensional. 
Such waves may propagate in every direction in an isotropic
compressible elastic material, but this is no longer the case for 
elastic anisotropic materials such as crystals. 
Possible directions of propagation of longitudinal homogeneous 
plane waves, called ``specific directions'' by Borgnis \cite{Borg55}
may be as few as three in an elastic anisotropic crystal. 
By assuming certain restrictions on the elastic constants, 
Hadamard \cite{Hada03} created a special model anisotropic elastic 
material in which longitudinal plane waves may propagate in every 
direction.

Here, we consider {\it inhomogeneous} plane waves. 
These waves are attenuated in a direction different from the 
propagation direction. 
They may be described in terms of bivectors \cite{BoHa93} --
complex vectors -- the amplitude bivector, $\mathbf{A}$ (say), 
and the slowness bivector, $\mathbf{S}$ (say), which may be written 
\cite{Haye84} $\mathbf{S} = N\mathbf{C}$, where the 
``directional bivector'' $\mathbf{C}$ is written 
$\mathbf{C} = m \mathbf{\hat{m}} + i \mathbf{\hat{n}}$ 
($\mathbf{\hat{m} \cdot \hat{n}}=0$, $m \ge 1$, 
$|\mathbf{\hat{m}}| = |\mathbf{\hat{n}}| = 1$) and $N$ is called 
the ``complex scalar slowness''. 
Once the directional bivector $\mathbf{C}$ is prescribed, 
the slowness $\mathbf{S}$ and amplitude $\mathbf{A}$ are determined 
from the equations of motion.
To prescribe $\mathbf{C}$ is equivalent to prescribing an ellipse 
with major semi-axis $m \mathbf{\hat{m}}$ and minor semi-axis 
$\mathbf{\hat{n}}$; 
this so-called ``directional ellipse'' for inhomogeneous plane waves 
is the analogue to the direction of propagation $\mathbf{n}$ for 
homogeneous plane waves \cite{Haye84}. 
The inhomogeneous plane wave is said to be ``longitudinal'' if 
$\mathbf{A}$ and $\mathbf{S}$ (and therefore also $\mathbf{C}$) are 
``parallel'': $\mathbf{A \wedge S} = \mathbf{0}$. 
What this means is that \cite{BoHa93} the ellipses of $\mathbf{A}$ 
and of $\mathbf{S}$ (and $\mathbf{C}$) are all ``parallel'', being 
similar -- same aspect ratio -- and being similarly situated 
-- parallel major axes and parallel minor axes. 
In particular we consider ``circularly polarized longitudinal 
inhomogeneous plane waves'' (CPLIPW). 
For such waves both $\mathbf{C}$ and $\mathbf{A}$ are isotropic, 
that is $\mathbf{C \cdot C} = 0$, $\mathbf{A \cdot A} = 0$; 
the ellipses corresponding to $\mathbf{C}$ and $\mathbf{A}$ are 
coplanar circles, the normal to the plane being $\mathbf{a}
 = \mathbf{\hat{m} \wedge \hat{n}}$. 

Here we seek to determine restrictions on the elastic constants such 
that for any choice of $\mathbf{a}$, the normal to the plane of 
$\mathbf{C}$ and $\mathbf{A}$ (which is the plane of the motion), 
a CPLIPW may propagate. 
We call the corresponding materials ``special''.

Starting with the constitutive equation for a general anisotropic 
elastic crystal (which involves twenty-one independent elastic 
constants), we obtain necessary and sufficient conditions on the 
constants in order that CPLIPWs may propagate in every plane. 
It turns out that nine linear relations among the elastic constants 
must be satisfied so that the special model material has at most 
twelve independent elastic constants. 
For such materials we determine the general structure of the 
(symmetric) acoustical tensor. 
The complex slowness $N$ of the CPLIPW is determined 
for all choices of isotropic $\mathbf{C}$. 
Because the wave is circularly polarized, one eigenvalue of the 
acoustical tensor is double \cite{Haye84}. 
The remaining simple eigenvalue corresponds to a ``transverse'' 
inhomogeneous plane wave that is transverse in the sense that 
its amplitude bivector $\mathbf{B}$ (say) is ``orthogonal'' to 
$\mathbf{C}$: $\mathbf{B \cdot C} = 0$, which means that the 
orthogonal projection of the ellipse of $\mathbf{B}$ upon the plane 
of $\mathbf{C}$ is a circle. 

The equation giving the complex slowness $N$ for the CPLIPW is of 
precisely the same structure as the equation giving the complex 
slowness of the transverse wave. 
Both are of the form $\mathbf{C \cdot \Theta C} = N^{-2}$, 
 $\mathbf{C \cdot C} = 0$, 
where $\mathbf{\Theta}$ is a real symmetric tensor. 
An ellipsoid $E$ (say) may be associated with $\mathbf{\Theta}$ 
\cite{BoHa93}. 
It is seen that the slowness bivectors are obtained by first 
determining the central ellipsoidal section $\mathcal{E}$ (say) 
of the ellipsoid $E$ by the central plane with normal $\mathbf{a}$. 
Then $\mathbf{\hat{m}}$ and $\mathbf{\hat{n}}$ are chosen to lie 
along the principal axes of $\mathcal{E}$. 
This fixes $\mathbf{C}$, and then $N$, and therefore $\mathbf{S}$
($= N\mathbf{C}$) is determined. 
For choices of $\mathbf{a}$ along the normals to the planes of 
central circular sections of the ellipsoid $E$, there is no 
propagating wave. 

Finally, we briefly consider the possibility of having CPLIPWs 
propagating in crystals of various classes: triclinic, 
monoclinic, etc. 
It is seen that CPLIPWs may not propagate in trigonal, tetragonal, 
and cubic crystals, nor in isotropic materials. 
It is seen that they may propagate in triclinic, monoclinic, 
orthorhombic, and 
hexagonal crystals, provided the linear relations 
among the elastic constants evoked above 
(or their specialization to those classes of symmetry)  are satisfied. 
As kindly pointed out by a referee, we must ensure that the crystals 
are purely elastic so that mechanical fields are not coupled to 
electrical fields. 
Hence in what follows, we restrict our attention to the following 
crystal classes: 
$\bar{1}$, 2/m, mmm, $\bar{3}$m, 4/m, 4/mmm, 6/m, 6/mmm, 432, m3, m3m.

\section{Basic equations}

The constitutive equations relating stress components $\sigma_{ij}$
with strain components $e_{ij}$ for a homogeneous anisotropic
elastic crystal are given by Hooke's law:
\begin{equation}
\sigma_{ij} = d_{ijkl} e_{kl}.
\end{equation}
Here $d_{ijkl}$ the elastic constants, or stiffnesses,  have the
symmetries,
\begin{equation} \label{symmetries}
d_{ijkl} = d_{jikl} = d_{klij},
\end{equation}
so that there are at most 21 independent stiffnesses.
Also,
\begin{equation}
2e_{ij} := \partial u_i / \partial x_j +  \partial u_j / \partial x_i,
\end{equation}
where $u_i$ are the displacement components: $u_i := x_i - X_i$.
Here $\mathbf{x}$ is the current position of a particle initially at
$\mathbf{X}$.

The equations of motion, in the absence of body forces, are given by
\begin{equation} \label{motion}
d_{ijkl} \partial^2 u_k / \partial x_l \partial x_j
  = \rho \partial^2 u_i / \partial t^2,
\end{equation}
where $\rho$ is the mass density of the crystal.

We consider displacements of the form
\begin{equation}  \label{u}
\mathbf{u} =
  \{\mathbf{A} \text{e}^{i \omega (\mathbf{S \cdot x} - t)} \}^+
     =       \text{e}^{- \omega \mathbf{S^- \cdot x}}
    \{\mathbf{A}^+ \cos \omega (\mathbf{S}^+ \mathbf{\cdot x} - t)
     - \mathbf{A}^- \sin \omega (\mathbf{S}^+ \mathbf{\cdot x} - t)\},
\end{equation}
where
$\mathbf{S} = \mathbf{S}^+ + i \mathbf{S}^-$ is the slowness bivector
(complex vector) \cite{Haye84, BoHa93} and
$\mathbf{A} = \mathbf{A}^+ + i \mathbf{A}^-$ is the amplitude bivector
(complex vector).
When $\mathbf{A}$ is parallel to $\mathbf{S}$, that is when
$\mathbf{A} = \alpha \mathbf{S}$, where $\alpha$ is some complex
number, the inhomogeneous wave is said to be ``longitudinal''
\cite{DeHa02, DeHa04}.
When $\mathbf{A}$ is isotropic, that is when $\mathbf{A \cdot A} = 0$,
the wave is circularly polarized, the ellipse associated with the
bivector $\mathbf{A}$ being a circle \cite{BoHa93}.

Circularly polarized longitudinal inhomogeneous waves are those waves
for which $\mathbf{A}$ and $\mathbf{S}$ are both isotropic and
parallel.
For such waves,
\begin{equation}
\mathbf{S \cdot S} = 0,
\end{equation}
or equivalently,
\begin{equation}
\mathbf{S}^+ \cdot \mathbf{S}^+
  = \mathbf{S}^- \cdot \mathbf{S}^-, \quad
   \mathbf{S}^+ \cdot \mathbf{S}^- = 0.
\end{equation}
Thus, the planes of constant phase are orthogonal to the planes of
constant amplitude; the waves propagate in the direction of
$\mathbf{S}^+$ whilst the amplitude decays in the direction of
$\mathbf{S}^-$.
The particle paths are circles in the plane of $\mathbf{A}^+$
and $\mathbf{A}^-$, or equivalently, because
$\mathbf{A} = \alpha \mathbf{S}$,
in the plane of $\mathbf{S}^+$ and $\mathbf{S}^-$.
The sense of description of the circle is from $\mathbf{S}^+$
towards $\mathbf{S}^-$, retrograde, similar to the sense of
Rayleigh waves propagating close to the free surface of a
semi-infinite isotropic elastic material.

Inserting \eqref{u} into \eqref{motion} gives the \textit{propagation
condition},
\begin{equation} \label{propagationCondition}
Q_{ik}(\mathbf{S}) A_k = \rho A_i, \quad
Q_{ik}(\mathbf{S}) = d_{ijkl} S_j S_l,
\end{equation}
where $Q_{ik}(\mathbf{S})$ is the \textit{acoustical tensor}
corresponding to the slowness bivector $\mathbf{S}$.
A systematic procedure for obtaining all solutions $\mathbf{A}$,
$\mathbf{S}$ of \eqref{propagationCondition} has been introduced
by Hayes \cite{Haye84} and is called the ``directional ellipse method''
or ``DE-method''.
It consists in writing $\mathbf{S}$ as
\begin{equation} \label{DE}
\mathbf{S} = N \mathbf{C},
\end{equation}
where $N$ is a complex number and $\mathbf{C}$ is any bivector of the
form
$\mathbf{C} = m \mathbf{\hat{m}} + i \mathbf{\hat{n}}$,
with $\mathbf{\hat{m}}$, $\mathbf{\hat{n}}$, two unit orthogonal
vectors and $m \ge 1$.
We call $N$ the \textit{complex scalar slowness} and $\mathbf{C}$ 
the \textit{directional bivector}.
Inserting \eqref{DE} into \eqref{propagationCondition} yields the
eigenvalue problem,
\begin{equation} \label{QA}
Q_{ik}(\mathbf{C})A_k = \rho N^{-2} A_i,
\end{equation}
for the complex symmetric tensor $Q_{ik}(\mathbf{C})$.
All solutions of the propagation condition may then be obtained
by prescribing $\mathbf{C}$ arbitrarily and solving \eqref{QA} for
$N^{-2}$ and $\mathbf{A}$, which gives the slowness bivector
through \eqref{DE}, and the amplitude bivector $\mathbf{A}$
(up to a complex scalar factor).
When, for some eigenvalue $N^{-2}$, the corresponding eigenbivector
$\mathbf{A}$ is parallel to the directional bivector $\mathbf{C}$,
the corresponding inhomogeneous plane wave is said to be longitudinal.
Thus, for longitudinal inhomogeneous plane waves the 
propagation condition becomes
\begin{equation} \label{QC}
Q_{ik}(\mathbf{C})C_k = \rho N^{-2} C_i,
\end{equation}
for some $N^{-2}$.
Hence, these waves are only possible
for directional bivectors $\mathbf{C}$ such that
$\mathbf{Q}(\mathbf{C})\mathbf{C}$
is parallel to $\mathbf{C}$, or equivalently,
\begin{equation}
\mathbf{C} \times \mathbf{Q}(\mathbf{C})\mathbf{C} = \mathbf{0},
\end{equation}
or
\begin{equation} \label{conditions}
\dfrac{Q_{1k}(\mathbf{C})C_k}{C_1} =
\dfrac{Q_{2k}(\mathbf{C})C_k}{C_2} =
\dfrac{Q_{3k}(\mathbf{C})C_k}{C_3}.
\end{equation}

Here we seek particular classes of materials which are such that
longitudinal inhomogeneous plane waves are possible
for any choice of $\mathbf{C}$ satisfying $\mathbf{C \cdot C} = 0$
(so that $\mathbf{C}$ is of the form 
$\mathbf{C} = \mathbf{\hat{m}} + i\mathbf{\hat{n}}$).
That is, we wish to determine under which conditions on the
stiffnesses $d_{ijkl}$ are CPLIPWs possible for all choices of 
the plane of $\mathbf{C}$,
or equivalently of the normal 
$\mathbf{a} = \mathbf{\hat{m} \times \hat{n}}$.

\section{Crystals for which longitudinal circularly polarized waves
are possible in all planes}

\subsection{Necessary and sufficient conditions}

Here we seek under which conditions on the stiffnesses $d_{ijkl}$,
we have:
$\mathbf{C} \times \mathbf{Q}(\mathbf{C})\mathbf{C} = \mathbf{0}$
for all $\mathbf{C} = \mathbf{\hat{m}} + i \mathbf{\hat{n}}$, or
equivalently, for all $\mathbf{C}$ satisfying
$\mathbf{C \cdot C} = 0$.

For convenience we adopt the Voigt \cite{Voig10} contracted notation
for the elastic stiffnesses,
\begin{equation}
d_{12} = d_{1122}, \quad d_{33} = d_{3333},  \quad
d_{45} = d_{2313}, \quad d_{66} = d_{1212},  \quad \text{etc.}
\end{equation}
With these notations,
\begin{align} \label{Q_ii}
& Q_{11}(\mathbf{C}) =
  d_{11}C_1^2 + d_{66}C_2^2 + d_{55}C_3^2
   +  2d_{16}C_1C_2 + 2d_{15}C_1C_3 + 2d_{56}C_2C_3,
\notag \\&
Q_{22}(\mathbf{C}) =
  d_{66}C_1^2 + d_{22}C_2^2 + d_{44}C_3^2
   +  2d_{26}C_1C_2 + 2d_{64}C_1C_3 + 2d_{24}C_2C_3,
\notag \\
& Q_{33}(\mathbf{C}) =
  d_{55}C_1^2 + d_{44}C_2^2 + d_{33}C_3^2
   +  2d_{45}C_1C_2 + 2d_{35}C_1C_3 + 2d_{34}C_2C_3,
  \end{align}
and
\begin{align} \label{Q_ij}
& Q_{12}(\mathbf{C}) =
  d_{16}C_1^2 + d_{26}C_2^2 + d_{45}C_3^2
\notag \\
& \phantom{0123456789}
   +  (d_{12}+d_{66})C_1C_2 + (d_{14}+d_{56})C_1C_3
     + (d_{25}+d_{46})C_2C_3,
\notag \\
&  Q_{23}(\mathbf{C}) =
  d_{56}C_1^2 + d_{24}C_2^2 + d_{34}C_3^2
\notag \\
& \phantom{0123456789}
   +  (d_{25}+d_{46})C_1C_2
     + (d_{36}+d_{45})C_1C_3
       + (d_{23}+d_{44})C_2C_3,
\notag \\
& Q_{31}(\mathbf{C}) =
  d_{15}C_1^2 + d_{46}C_2^2 + d_{35}C_3^2
\notag \\
& \phantom{0123456789}
   +  (d_{14}+d_{56})C_1C_2
     + (d_{13}+d_{55})C_1C_3
       + (d_{36}+d_{45})C_2C_3.
  \end{align}

Consider first $\mathbf{C} = (0,1,i)$. It is isotropic.
For this $\mathbf{C}$, conditions \eqref{conditions} become
$Q_{1k}C_k = 0$ and $iQ_{2k}C_k = Q_{3k}C_k$,
which read, explicitly,
\begin{align}
& (d_{36} + 2d_{45} - d_{26}) - i(d_{25} + 2d_{46} - d_{35})=0,
\notag \\
& 4(d_{34} - d_{24}) + i(d_{22} + d_{33} - 2d_{23} - 4d_{44}) = 0.
\end{align}
Hence,
\begin{equation} \label{cond1}
d_{26} = d_{36} + 2d_{45}, \quad
d_{35} = d_{25} + 2d_{46}, \quad
d_{24} = d_{34}, \quad
4d_{44} = d_{22} + d_{33} - 2d_{23}.
\end{equation}

We then consider in turn  $\mathbf{C} = (i,0,1)$ and
$\mathbf{C} = (1,i,0)$.
These choices yield conditions of the type \eqref{cond1}, which may be
read off from \eqref{cond1} on cycling the indices
$1 \rightarrow 2 \rightarrow 3 \rightarrow 1$,
$4 \rightarrow 5 \rightarrow 6 \rightarrow 4$.
The complete set of conditions obtained in this way is
\begin{align} \label{cond2}
& d_{16} = d_{26} = d_{36} + 2d_{45}, \quad
d_{35} = d_{15} = d_{25} + 2d_{46}, \quad
d_{24} = d_{34} = d_{14} + 2d_{56},
\nonumber \\
& 4d_{44} = d_{22} + d_{33} - 2d_{23}, \;
    4d_{55} = d_{33} + d_{11} - 2d_{13}, \;
    4d_{66} = d_{11} + d_{22} - 2d_{12}.
\end{align}
We refer to materials whose stiffnesses satisfy 
these conditions as ``special''.  
It follows that ``special'' materials have at most twelve 
independent elastic stiffnesses.
For instance, they are 
\begin{equation} \label{constitutive}
\begin{array}{c c c c c c}
 d_{11} & d_{12} & d_{13} & d_{14}  & d_{15}  & d_{16}  \\
        & d_{22} & d_{23} & d_{24}  & d_{25}  & \bullet \\
        &        & d_{33} & \bullet & \bullet & d_{36}  \\
        &        &        & \bullet & \bullet & \bullet \\
        &        &        &         & \bullet & \bullet \\
        &        &        &         &         & \bullet
\end{array}
\end{equation}
and the remaining elastic constants (denoted by ``$\bullet$''
above) are determined from those 12 as
\begin{align}
&  d_{34} = d_{24}, 
&& d_{44}  = \quart (d_{22} + d_{33} - 2d_{23}), 
&&  d_{45}  = \halft (d_{26} - d_{36}),
\notag \\
&  d_{35} = d_{15},
&& d_{55} = \quart (d_{11} + d_{33} - 2d_{13}), 
&& d_{56}  = \halft (d_{34} - d_{14})
\notag \\
&  d_{26} = d_{16},
&& d_{66}  = \quart (d_{11} + d_{22} - 2d_{12}),
&& d_{46}  = \halft (d_{35} - d_{25}).
\end{align}
We note that out of the nine conditions \eqref{cond2}, 
six are ``structurally invariant'' \cite{Ting00} for some
rotations of the coordinate system.
In particular,  if the two conditions
\begin{equation}
  d_{16} - d_{26} = d_{11} + d_{22} - 2d_{12} -4d_{66} = 0,
\end{equation}
are satisfied in the coordinate system linked
to the crystallographic axes ($O x_1 x_2 x_3$), then they are
also satisfied by the stiffnesses $d^*_{ij}$ obtained from the
$d_{ij}$ after any rotation of the coordinate system about the
$x_3$ axis.
These invariants are Type 1B in Ting's classification \cite{Ting00}.
Similarly, the conditions
\begin{equation}
  d_{15} - d_{35} = d_{11} + d_{33} - 2d_{13} -4d_{55} = 0,
\end{equation}
are invariants under rotation of the coordinate system about the
$x_2$ axis, and
\begin{equation}
  d_{24} - d_{34} = d_{22} + d_{33} - 2d_{23} -4d_{44} = 0,
\end{equation}
are invariants after rotation of the coordinate system about the
$x_1$ axis.
 
When the relations \eqref{cond2} hold, it may be checked
by direct calculation, using \eqref{Q_ii}, \eqref{Q_ij}, and taking
into account the relation $C_1^2+C_2^2+C_3^2=0$, that
\begin{equation}
Q_{ik}(\mathbf{C})C_k = \rho N_L^{-2}(\mathbf{C}) C_i,
\end{equation}
for all isotropic bivectors $\mathbf{C}$,
with $N_L^{-2}(\mathbf{C})$ given by
\begin{equation} \label{N_L}
\rho N_L^{-2}(\mathbf{C})
   = \halft (d_{11}C_1^2 + d_{22}C_2^2 + d_{33}C_3^2)
     +2(d_{16}C_1C_2 + d_{35}C_1C_3 + d_{24}C_2C_3).
\end{equation}
It follows that the relations \eqref{cond2} between the elastic
stiffnesses are the necessary and sufficient conditions for
CPLIPWs to
propagate for all choices of isotropic directional bivectors, or,
equivalently, for all choices of the polarization plane.

The expression \eqref{N_L} may also be written
\begin{equation}
\rho N_L^{-2}(\mathbf{C})
   = \mathbf{C} \cdot \mathbf{\Psi}_L \mathbf{C},
\end{equation}
where $\mathbf{\Psi}_L$ is given by
\begin{equation}  \label{Psi_L}
\mathbf{\Psi}_L =   \begin{bmatrix}
     \halft d_{11} & d_{16}          &  d_{35}       \\
     d_{16}        & \halft d_{22}   &  d_{24}       \\
     d_{35}        & d_{24}          &  \halft d_{33}
    \end{bmatrix}.
\end{equation}
Associated with $\mathbf{\Psi}_L$ is the
``$\mathbf{\Psi}_L$-ellipsoid'':
$\mathbf{x} \cdot (\mathbf{\Psi}_L + p\mathbf{1})\mathbf{x} = 1$,
where $p$ is chosen such that $ (\mathbf{\Psi}_L + p\mathbf{1})$ is
positive definite.
If $\mathbf{C}$ is chosen to lie on either plane of central circular
section of the $\mathbf{\Psi}_L$-ellipsoid, then
$\mathbf{C} \cdot \mathbf{\Psi}_L \mathbf{C} = 0$ and thus
$N_L^{-2}(\mathbf{C}) = 0$: the corresponding waves do not propagate.

In the next section, we derive the general structure of the acoustical
tensor for ``special'' materials.

\subsection{Acoustical tensor}

 From the previous section, we know that the acoustical tensor
$Q_{ik}({\bf C})$, with ${\bf C}\cdot {\bf C}=0$, of
``special'' materials admits
$\rho N_L^{-2}(\mathbf{C})$ given by \eqref{N_L} as an eigenvalue.
The corresponding eigenvector $\mathbf{C}$ is isotropic, and so
\cite{Haye84} this eigenvalue is a double eigenvalue.
It follows that the other (simple) eigenvalue
$\rho N_*^{-2}(\mathbf{C})$, say, is given by
$\rho N_*^{-2}(\mathbf{C})
  = \text{tr }\mathbf{Q}(\mathbf{C}) - 2 \rho N_L^{-2}(\mathbf{C})$,
that is,
\begin{multline} \label{N_*}
\rho N_*^{-2}(\mathbf{C}) =
  (d_{55}+d_{66})C_1^2 + (d_{44}+d_{66})C_2^2 + (d_{44}+d_{55})C_3^2
  \\ +2d_{45}C_1C_2 + 2d_{46}C_1C_3 + 2d_{56}C_2C_3.
\end{multline}
The expression \eqref{N_*} may also be written
\begin{equation} \label{N_*Psi_T}
\rho N_*^{-2}(\mathbf{C})
   = \mathbf{C} \cdot \mathbf{\Psi}_T \mathbf{C},
\end{equation}
where $\mathbf{\Psi}_T$ is given by
\begin{equation}  \label{Psi_T}
\mathbf{\Psi}_T =  \begin{bmatrix}
      d_{55}+d_{66} &   d_{45}           &    d_{46}       \\
      d_{45}        & d_{44}+d_{66}      &    d_{56}       \\
      d_{46}        &   d_{56}           &  d_{44} + d_{55}
                           \end{bmatrix}.
\end{equation}
Associated with $\mathbf{\Psi}_T$ is the
``$\mathbf{\Psi}_T$-ellipsoid'':
$\mathbf{x} \cdot (\mathbf{\Psi}_T + p\mathbf{1})\mathbf{x} = 1$,
where $p$ is chosen such that $ (\mathbf{\Psi}_T + p\mathbf{1})$ is
positive definite.
If $\mathbf{C}$ is chosen to lie on either plane of central circular
section of the $\mathbf{\Psi}_T$-ellipsoid, then
$\mathbf{C} \cdot \mathbf{\Psi}_T \mathbf{C} = 0$ and so
$N_*^{-2}(\mathbf{C}) = 0$: there is no corresponding 
transverse circularly polarized propagating wave.

Now we compute the components of the matrix
$\mathbf{\Gamma}(\mathbf{C}):=
   \mathbf{Q}(\mathbf{C})-\rho N_*^{-2}(\mathbf{C})\mathbf{1}$.
We find, using the conditions \eqref{cond2} and
$C_1^2+C_2^2+C_3^2=0$, that
\begin{align}
& \Gamma_{11}(\mathbf{C}) =
  (\mu + \nu - \lambda)C_1^2 + 2\gamma C_1 C_2 + 2\beta C_1 C_3,
\notag \\
& \Gamma_{22}(\mathbf{C}) =
  (\nu + \lambda - \mu)C_2^2 + 2\alpha C_2 C_3 + 2\gamma C_1 C_2,
\notag \\
& \Gamma_{33}(\mathbf{C}) =
  (\lambda + \mu - \nu)C_3^2 + 2\beta C_1 C_3 + 2\alpha C_2 C_3,
\end{align}
and that
\begin{align}
& \Gamma_{12}(\mathbf{C}) =
  - \gamma C_3^2 + \nu C_1 C_2 + \alpha C_1 C_3 + \beta C_2 C_3,
\notag \\
& \Gamma_{23}(\mathbf{C}) =
  -\alpha C_1^2 + \lambda C_2 C_3 + \beta C_1 C_2 + \gamma C_1 C_3,
\notag \\
& \Gamma_{31}(\mathbf{C}) =
   - \beta C_2^2 + \mu C_1 C_3 + \gamma C_2 C_3 + \alpha C_1 C_2,
\end{align}
where
\begin{align}
&     \lambda:= \halft (d_{22}+d_{33}-2d_{44}),
&&   \mu:= \halft (d_{11}+d_{33}-2d_{55}),
&&   \nu:= \halft (d_{11}+d_{22}-2d_{66}),
\notag \\
&   \alpha:= d_{14}+d_{56},
&& \beta:= d_{25}+d_{46},
&& \gamma:= d_{36}+d_{45}.
\end{align}
Let $\mathbf{M}$ be the real symmetric matrix defined as
\begin{equation}
\mathbf{M}  := \begin{bmatrix}
      \lambda   &  -\gamma   &  -\beta   \\
      -\gamma  &   \mu          &  -\alpha \\
      -\beta       &  -\alpha      &   \nu
                           \end{bmatrix}.
\end{equation}
It can be checked that
\begin{equation}
\mathbf{\Gamma}(\mathbf{C}) =
  (\lambda + \mu + \nu) \mathbf{C} \otimes \mathbf{C}
    - \mathbf{M C} \otimes \mathbf{C}
     - \mathbf{C} \otimes \mathbf{M C}.
\end{equation}
Alternatively, introducing
$\mathbf{\hat{M}} :=
   \halft (\lambda + \mu + \nu)\mathbf{1} - \mathbf{M}$,
the matrix $\mathbf{\Gamma}(\mathbf{C})$ may be written as
\begin{equation}
\mathbf{\Gamma}(\mathbf{C}) =
  \mathbf{\hat{M} C} \otimes \mathbf{C}
     + \mathbf{C} \otimes \mathbf{\hat{M} C},
\end{equation}
where, explicitly,
{\small
\begin{equation} \label{Mhat}
\mathbf{\hat{M}}  = \begin{bmatrix}
  \halft (d_{11}+d_{44}-d_{55}-d_{66}) & d_{36}+d_{45} & d_{25}+d_{46}
\\
  d_{36}+d_{45} & \halft (d_{22}+d_{55}-d_{44}-d_{66}) & d_{14}+d_{56}
\\
  d_{25}+d_{46} & d_{14}+d_{56} & \halft (d_{33}+d_{66}-d_{44}-d_{55})
                           \end{bmatrix}.
\end{equation}
}
Hence, noting that $\mathbf{C \cdot C} =0$ and recalling the
definition of $\mathbf{\Gamma}(\mathbf{C})$,
we find that the acoustical tensor may be put in the form,
\begin{equation} \label{decompQ}
\mathbf{Q}(\mathbf{C}) =
  \rho N_*^{-2}(\mathbf{C}) \mathbf{1} + \mathbf{\hat{M} C}
   \otimes \mathbf{C}
     + \mathbf{C} \otimes \mathbf{\hat{M} C}.
\end{equation}
This decomposition of the acoustical tensor
shows directly that the isotropic bivector $\mathbf{C}$ is an
eigenvector of $\mathbf{Q}(\mathbf{C})$, whose eigenvalue
$\rho N_L^{-2}(\mathbf{C})$ given by \eqref{N_L}, can equivalently
be written
\begin{equation} \label{N_L/N_*}
\rho N_L^{-2}(\mathbf{C})
    = \rho N_*^{-2}(\mathbf{C}) + \mathbf{C \cdot \hat{M}C}.
\end{equation}
Also, the eigenvector corresponding to the eigenvalue
$\rho N_*^{-2}(\mathbf{C})$ is $\mathbf{C \times \hat{M}C}$.
We call the corresponding wave the ``transverse'' inhomogeneous plane
wave, because its amplitude bivector ${\bf A}$ is
orthogonal to ${\bf C}$: ${\bf C}\cdot {\bf A}=0$.
In general it is elliptically polarized because
$(\mathbf{C \times \hat{M}C}) \cdot (\mathbf{C \times \hat{M}C})
= - (\mathbf{C \cdot \hat{M}C})^2 \ne 0$.
Of course $\mathbf{C} \cdot
              [\mathbf{C} \times \hat{\mathbf{M}}\mathbf{C}] = 0$,
which means \cite{Haye84} that the orthogonal projection,
upon the plane of $\mathbf{C}$, of the ellipse associated with
the amplitude bivector $\mathbf{C \times \hat{M}C}$, is a circle.

Let $\mathbf{\tilde{C}}$ be a choice of $\mathbf{C}$ for which 
$\mathbf{\tilde{C} \cdot \hat{M}\tilde{C}} 
  = \mathbf{\tilde{C} \cdot \tilde{C}} = 0$, so that 
$\mathbf{\tilde{C}}$ is parallel to 
$\mathbf{\tilde{C} \times \hat{M}\tilde{C}}$ and there is
a triple eigenvalue $\rho N_L^{-2}(\mathbf{\tilde{C}})
                         = \rho N_*^{-2}(\mathbf{\tilde{C}})$.
This special case occurs when the plane of $\mathbf{\tilde{C}}$ is
one of the two planes of central circular section of the
the ``$\mathbf{\hat{M}}$-ellipsoid'':
$\mathbf{x} \cdot (\mathbf{\hat{M}} + p\mathbf{1})\mathbf{x} = 1$,
where $p$ is chosen such that $ (\mathbf{\hat{M}} + p\mathbf{1})$ is
positive definite.

\subsubsection*{Remark: A relationship between
the scalar slownesses of certain waves}

The form \eqref{N_L/N_*} of the relation triggers the following
remark.
Let the matrix $\mathbf{\hat{M}}$ have eigenvalues
$p_1$, $p_2$, $p_3$ and corresponding unit orthogonal eigenvectors
$\mathbf{a}_1$, $\mathbf{a}_2$, $\mathbf{a}_3$ so that 
\begin{equation}
\mathbf{\hat{M}} =
p_1 \mathbf{a}_1 \otimes \mathbf{a}_1 +
  p_2 \mathbf{a}_2 \otimes \mathbf{a}_2 +
   p_3 \mathbf{a}_3 \otimes \mathbf{a}_3,
\quad
\mathbf{a}_i \cdot \mathbf{a}_j = \delta_{ij}.
\end{equation}
Then consider the following isotropic bivectors $\mathbf{C}_1$,
$\mathbf{C}_2$, $\mathbf{C}_3$,
\begin{equation} \label{Ci}
\mathbf{C}_1 := \mathbf{a}_2 + i \mathbf{a}_3, \quad
  \mathbf{C}_2 := \mathbf{a}_3 + i \mathbf{a}_1, \quad
   \mathbf{C}_3 := \mathbf{a}_1 + i \mathbf{a}_2.
\end{equation}
Clearly,
\begin{equation}
  \mathbf{C}_1 \mathbf{\cdot \hat{M} C}_1 = p_2 - p_3, \quad
   \mathbf{C}_2 \mathbf{\cdot \hat{M} C}_2 = p_3 - p_1, \quad
    \mathbf{C}_3 \mathbf{\cdot \hat{M} C}_3 = p_1 - p_2,
\end{equation}
so that
\begin{equation} \label{CiSum}
\sum_{i=1}^3  \mathbf{C}_i \mathbf{\cdot \hat{M} C}_i = 0.
\end{equation}
Then, from \eqref{N_L/N_*} and \eqref{CiSum}, we have the relation
\begin{equation}
\sum_{i=1}^3  N_L^{-2} (\mathbf{C}_i) =
  \sum_{i=1}^3  N_*^{-2} (\mathbf{C}_i),
\end{equation}
for the $\mathbf{C}_i$ given by \eqref{Ci}.

\subsubsection*{Example}

To illustrate the results of this section, we work out a
simple example.
Let $\mathbf{C} = \mathbf{i} + i \mathbf{j}$.
Then the corresponding acoustical tensor
$\mathbf{Q}(\mathbf{i} + i \mathbf{j}) $ is
{\small
\begin{equation}
                        \begin{bmatrix}
  d_{11}-d_{66}+2id_{16} & i(d_{66}+d_{12})
                         & d_{15}-d_{46} +i(d_{56}+d_{14}) \\
  i(d_{66}+d_{12})       & d_{66}-d_{22}+2id_{26}
                         &  -(d_{56}+d_{14}) +i(d_{15}-d_{46})   \\
  d_{15}-d_{46} +i(d_{56}+d_{14}) & -(d_{56}+d_{14}) +i(d_{15}-d_{46})
                                  &  d_{55}-d_{44} + 2id_{45}
                           \end{bmatrix}.
\end{equation}
}
Computing $\mathbf{Q}(\mathbf{C})\mathbf{C}$ and using the
conditions \eqref{cond2}, we find that $\mathbf{C}$ is
indeed an eigenvector of the acoustical tensor, with eigenvalue
given by \eqref{N_L},
\begin{equation}
\mathbf{Q}(\mathbf{C})\mathbf{C}
   = \rho N_L^{-2} \mathbf{C},
\quad
\rho N_L^{-2} = \halft (d_{11} - d_{22}) + 2id_{16}.
\end{equation}
Further, computing $\mathbf{C}\times \mathbf{\hat{M}C}$ and then
$\mathbf{Q}(\mathbf{C})(\mathbf{C}\times \mathbf{\hat{M}C})$ and using
the conditions \eqref{cond2}, we find that
$\mathbf{C}\times \mathbf{\hat{M}C}$ is
also an eigenvector of the acoustical tensor, with eigenvalue
now given by \eqref{N_*},
\begin{equation}
\mathbf{Q}(\mathbf{C})(\mathbf{C}\times \mathbf{\hat{M}C})
  = \rho N_*^{-2} (\mathbf{C}\times \mathbf{\hat{M}C}),
\quad
\rho N_*^{-2} = d_{55} - d_{44} + 2id_{45}.
\end{equation}

In this simple example, we prescribed the normal to the plane of the
isotropic slowness bivector (or equivalently, to the plane of the
amplitude bivector) to be $\mathbf{k}$.
Then, we chose $\mathbf{C}$ to be $\mathbf{i} + i \mathbf{j}$,
leading to a complex eigenvalue.
In the next section, we show that it is always possible to choose
$\mathbf{C}$ such that the corresponding eigenvalue is a real positive
number.

\subsubsection*{Remark: General form of the acoustical tensor for the 
``special'' materials}

Using the relations \eqref{cond2} in the expressions 
$Q_{ij}(\mathbf{C})$ given by  \eqref{Q_ii} and \eqref{Q_ij}, 
it may be seen that without any restrictions on $\mathbf{C}$, the 
acoustical tensor $\mathbf{Q}(\mathbf{C})$ for the ``special'' 
materials may be written as
\begin{equation} \label{newQ}
  \mathbf{Q}(\mathbf{C}) = \rho N_*^{-2}(\mathbf{C}) \mathbf{1} 
 + \mathbf{\hat{M}C} \otimes \mathbf{C}
  + \mathbf{C}\otimes  \mathbf{\hat{M}C}
   + (\mathbf{C \cdot C}) \mathbf{\Delta}.
\end{equation}
Here the expression for $\rho N_*^{-2}(\mathbf{C})$ is given by 
\eqref{N_*} and $ \mathbf{\Delta}$ is defined by 
\begin{equation} \label{newDelta}
  \mathbf{\Delta} =       \begin{bmatrix}
      -d_{44} &  d_{45} &  d_{46} \\
       d_{45} & -d_{55} &  d_{56} \\
       d_{46} &  d_{56} & -d_{66}
                           \end{bmatrix}.
\end{equation}

Comparing  \eqref{Psi_T} and \eqref{newDelta}, we note that
$\mathbf{\Psi}_T 
  = \mathbf{\Delta} - (\text{tr }\mathbf{\Delta}) \mathbf{1}$, 
so that by  \eqref{N_*Psi_T},
\begin{equation}
   \rho N_*^{-2}(\mathbf{C}) = \mathbf{C \cdot \Delta C} 
     - (\text{tr }\mathbf{\Delta})\mathbf{C \cdot C}. 
\end{equation}
It follows that 
\begin{equation} \label{QQ}
  \mathbf{Q}(\mathbf{C}) = 
    [\mathbf{C \cdot \Delta C} 
      - (\text{tr }\mathbf{\Delta})\mathbf{C \cdot C}] \mathbf{1} 
        + \mathbf{\hat{M}C} \otimes \mathbf{C}
          + \mathbf{C}\otimes  \mathbf{\hat{M}C}
             + (\mathbf{C \cdot C}) \mathbf{\Delta},
\end{equation}
which is an expression for the acoustical tensor 
$\mathbf{Q}(\mathbf{C})$ for ``special'' materials, given in terms of 
only two matrices, $\mathbf{\Delta}$ and  $\mathbf{\hat{M}}$.

We note for ``special'' materials that $d_{ijkl}$ may be written 
\begin{multline}
  d_{ijkl} = \Delta_{jl}\delta_{ik} + \Delta_{ik}\delta_{jl} 
    + \Delta_{il}\delta_{jk} +  \Delta_{jk}\delta_{il} 
      - \Delta_{ij}\delta_{kl} - \Delta_{kl}\delta_{ij} \\
        - (\text{tr }\mathbf{\Delta})
           (\delta_{ik}\delta_{jl} +  \delta_{jk}\delta_{il} 
              - \delta_{ij}\delta_{kl}) 
           + \hat{M}_{ij}\delta_{kl}  + \hat{M}_{kl}\delta_{ij}.  
\end{multline}

\subsubsection*{Remark: Homogeneous plane waves in ``special'' 
materials}

For \textit{homogeneous} plane waves propagating in the direction 
$\mathbf{n}$ in the ``special'' materials, let 
$\mathbf{u} = \mathbf{A} \exp i\omega(S\mathbf{n \cdot x}-t)$. 
The corresponding propagation condition is then 
\begin{equation}
  \mathbf{Q}(\mathbf{n}) \mathbf{A} = \rho S^2 \mathbf{A},
\end{equation}
where $  \mathbf{Q}(\mathbf{n})$ is given by \eqref{QQ} with $n_1$, 
$n_2$, $n_3$ replacing $C_1$, $C_2$, $C_3$, respectively. 
Thus
\begin{equation}
 \mathbf{Q}(\mathbf{n}) = 
    [\mathbf{n \cdot \Delta n} 
      - (\text{tr }\mathbf{\Delta})] \mathbf{1} 
        + \mathbf{\hat{M}n} \otimes \mathbf{n}
          + \mathbf{n}\otimes  \mathbf{\hat{M}n}
             +  \mathbf{\Delta}. 
\end{equation}

We note a special solution.

Let the Hamiltonian decomposition of $\mathbf{\Delta}$ be given by 
\cite{BoHa93}
\begin{equation}
  \mathbf{\Delta} = \mu \mathbf{1} 
    + \kappa (\mathbf{h^+} \otimes \mathbf{h^-}
                + \mathbf{h^-} \otimes \mathbf{h^+}),
\end{equation}
where $\mu$, $\kappa$ are constants and $\mathbf{h^\pm}$ are constant 
unit vectors. 
Choose $\mathbf{n} = \mathbf{n^*}$ such that 
\begin{equation}
  (\mathbf{n^*} \times \mathbf{\hat{M} n^*}) \times 
    (\mathbf{h^+} \times \mathbf{h^-}) = \mathbf{0}.
\end{equation}
(It is always possible to do this.)
A solution for a homogeneous plane wave propagating along 
$\mathbf{n^*}$ is 
\begin{equation}
 \mathbf{u} =  
  (\mathbf{h^+} \times \mathbf{h^-}) 
    \exp i\omega(N^{-1} \mathbf{n^* \cdot x}-t), 
\quad 
  \text{where} \quad 
    \rho N^{-2} = \mu 
                    + \mathbf{n^* \cdot \Delta n^*} 
                      - (\text{tr }\mathbf{\Delta}).
\end{equation}

\section{Description of the waves}

Before proceeding, we note that the two equations \eqref{N_L} and
\eqref{N_*}, giving the complex scalar slowness of the longitudinal
circularly polarized wave and of the transverse elliptically polarized
wave, respectively, have the same form,
\begin{equation} \label{CpsiC}
\mathbf{C \cdot \Psi C} = \rho N^{-2}, \quad
\mathbf{C \cdot C} =0,
\end{equation}
where $\mathbf{\Psi}$ is a real symmetric
tensor given by $\mathbf{\Psi} = \mathbf{\Psi}_L$
(see \eqref{Psi_L}) for the longitudinal wave and by
$\mathbf{\Psi} = \mathbf{\Psi}_T$
(see \eqref{Psi_T}) for the transverse wave.
Accordingly, we determine the details of the slowness of the
corresponding wave solutions.
The results may be adapted either to the longitudinal wave
or to the transverse wave by replacing $\mathbf{\Psi}$ with
either $\mathbf{\Psi}_L$ or $\mathbf{\Psi}_T$.

\subsection{Construction of the slowness bivector}

Let $\Psi_1$, $\Psi_2$, $\Psi_3$ be the eigenvalues of
  $\mathbf{\Psi}$ and $\mathbf{e}_1$, $\mathbf{e}_2$, $\mathbf{e}_3$
the corresponding orthogonal unit eigenvectors.
We assume that the eigenvalues
are ordered as $\Psi_1>\Psi_2>\Psi_3$.
We note that the pair \eqref{CpsiC} is equivalent to the pair
\begin{equation} \label{CpsiCp}
\mathbf{C} \cdot (\mathbf{\Psi} + p \mathbf{1}) \mathbf{C}
= \rho N^{-2}, \quad
\mathbf{C \cdot C} =0,
\end{equation}
where $p$ is an arbitrary constant.
By choosing $p$ suitably large and positive, so that $\Psi_3 + p>0$,
we may define the positive definite matrix
$(\mathbf{\Psi} + p\mathbf{1})$ and associate the
ellipsoid
\begin{equation}
\mathbf{x} \cdot (\mathbf{\Psi} + p\mathbf{1}) \mathbf{x}
   = 1,
\end{equation}
with $\mathbf{\Psi}$.
We call this the ``$\mathbf{\Psi}$-ellipsoid''.
The planes of central circular section of the
$\mathbf{\Psi}$-ellipsoid have unit normals $\mathbf{h^\pm}$ given by
\cite{BoHa93}
\begin{equation} \label{optic}
(\Psi_1 - \Psi_3)^\halft \mathbf{h^\pm}
  = (\Psi_1 - \Psi_2)^\halft \mathbf{e}_1
        \pm (\Psi_2 - \Psi_3)^\halft \mathbf{e}_3.
\end{equation}
We call these normals $\mathbf{h^\pm}$, the ``optic axes'' of the
$\mathbf{\Psi}$-ellipsoid and note that they are independent of $p$.

To describe the slownesses of the waves corresponding to \eqref{CpsiC},
we recall that
\begin{equation} \label{S,C}
\mathbf{S} = N \mathbf{C}, \quad
\mathbf{C} = \mathbf{\hat{m}} + i \mathbf{\hat{n}}, \quad
\mathbf{\hat{m} \cdot \hat{m}} = \mathbf{\hat{n} \cdot \hat{n}} =1,
\quad
\mathbf{\hat{m} \cdot \hat{n}} =0.
\end{equation}
Here $\mathbf{C}$ is such that $\mathbf{C \cdot C} = 0$ and thus
$\mathbf{S \cdot S} = 0$.
Also ($\mathbf{\hat{m}}, \mathbf{\hat{n}}$) is \textit{any} pair of
orthogonal unit vectors in the plane of $\mathbf{S}$, or equivalently
in the plane with unit normal
$\mathbf{a} := \mathbf{\hat{m}} \times \mathbf{\hat{n}}$.
In \eqref{S,C}, $N$ is a scalar to be determined from the equations
\eqref{CpsiC}.

Because we are at liberty to choose for
($\mathbf{\hat{m}}, \mathbf{\hat{n}}$)
any orthogonal unit pair in the plane with normal $\mathbf{a}$,
we choose ($\mathbf{\hat{m}}, \mathbf{\hat{n}}$) along the principal
axes of the elliptical section of the $\mathbf{\Psi}$-ellipsoid by the
central plane $\mathbf{a \cdot x}=0$.
Specifically, we take  $\mathbf{\hat{m}}$ as the unit vector along the
minor axis of this ellipse and  $\mathbf{\hat{n}}$ along the major
axis.
In that event,
\begin{equation} \label{mPsin}
  \mathbf{\hat{m} \cdot \Psi \hat{n}} =0,
\end{equation}
and using \eqref{S,C} and \eqref{mPsin}, the pair \eqref{CpsiC} give
\begin{equation} \label{Nmn}
\rho N^{-2} =
  \mathbf{\hat{m} \cdot \Psi \hat{m}}
      - \mathbf{\hat{n} \cdot \Psi \hat{n}} > 0,
\end{equation}
in general, so that $N^{-1}$ is purely real: $N^{-1} = v$, say.
Also,
\begin{equation} \label{S+-}
  \mathbf{S}^+ = v^{-1} \mathbf{\hat{m}}, \quad
  \mathbf{S}^- = v^{-1} \mathbf{\hat{n}},
\end{equation}
so that the planes of constant phase (amplitude) are:
$\mathbf{\hat{m} \cdot x}=$ constant
($\mathbf{\hat{n} \cdot x}=$ constant).

Of course, if
$ \mathbf{\hat{m} \cdot \Psi \hat{m}}
       = \mathbf{\hat{n} \cdot \Psi \hat{n}}$,
so that the radii to the $\mathbf{\Psi}$-ellipsoid along the
orthogonal unit vectors are equal, then the plane
$\mathbf{a \cdot x}=0$ is a plane of central circular section of the
$\mathbf{\Psi}$-ellipsoid, and from  \eqref{Nmn} there is no
propagating solution: $\rho N^{-2} = 0$.

Thus, in general, the slowness bivectors corresponding to
\eqref{CpsiC} are obtained by first
determining the central elliptical section of the
$\mathbf{\Psi}$-ellipsoid by the plane $\mathbf{a \cdot x}=0$.
Then $\mathbf{\hat{m}}$ and $\mathbf{\hat{n}}$ are chosen along the
principal axes of the ellipse, and $\mathbf{S}^+$ and $\mathbf{S}^-$
are given by \eqref{Nmn} \eqref{S+-}.

To complete the picture we recall the results \cite{BoHa93}
for the determination of the principal axes of the central elliptical
section of the $\mathbf{\Psi}$-ellipsoid by the plane
$\mathbf{a\cdot x}=0$.

In the determination there are three cases to be considered:

Case(i) Normal $\mathbf{a}$ not coplanar with the optic axes.

Case(ii) Normal $\mathbf{a}$ coplanar with the optic axes but not
  parallel to either optic axis.

Case(iii) Normal $\mathbf{a}$ parallel to an optic axis.

We can dispose of Case (iii) immediately because we have just seen
that there is no propagating circularly polarized solution when
$\mathbf{a \cdot x}=0$ is a plane of central circular section of the
$\mathbf{\Psi}$-ellipsoid.

In Case (i), $\mathbf{a \cdot h^+ \times h^-} \ne 0$,
it has been shown \cite{BoHa93} that $\mathbf{\hat{m}}$,
$\mathbf{\hat{n}}$ are in the direction of $\mathbf{r^\pm}$ given by
\begin{equation} \label{mn1}
\mathbf{r^\pm} =
  [\mathbf{h^+} - (\cos \phi^+)\mathbf{a}]/(\sin \phi^+)
   \pm
  [\mathbf{h^-} - (\cos \phi^-)\mathbf{a}]/(\sin \phi^-),
\end{equation}
where $\phi^\pm$ is the angle between $\mathbf{a}$ and the optic axis
$\mathbf{h^\pm}$.
Essentially, $\mathbf{\hat{m}}$ and $\mathbf{\hat{n}}$ are along the
internal and external bisectors of the angle between the orthogonal
projections of $\mathbf{h^+}$ and $\mathbf{h^-}$ onto the plane
$\mathbf{a \cdot x}=0$ (this is the Fresnel construction of Optics.)

In Case (ii), $\mathbf{a \cdot h^+ \times h^-} = 0$,
$\mathbf{a  \times h^\pm} \ne \mathbf{0}$,
it has been shown \cite{BoHa93} that
\begin{equation} \label{mn2}
\mathbf{\hat{m}} =
  [\mathbf{h^+} - (\cos \phi^+)\mathbf{a}]/(\sin \phi^+),
   \quad
  \mathbf{\hat{n}} = (\mathbf{h^+} \times \mathbf{a})/(\sin \phi^+).
\end{equation}
Essentially, $\mathbf{\hat{m}}$ is along the orthogonal projection of
$\mathbf{h^+}$ (or $\mathbf{h^-}$) onto the plane
$\mathbf{a \cdot x}=0$, and $\mathbf{\hat{n}}$ is orthogonal to that
plane.

This means that as the direction of the unit normal $\mathbf{a}$ is
varied, the corresponding $\mathbf{\hat{m}}$ and $\mathbf{\hat{n}}$
are given by \eqref{mn1} and \eqref{mn2}.
Also, as shown by Boulanger and Hayes \cite{BoHa93}, in Cases
(i)  and (ii),
$\mathbf{\hat{m}\cdot\Psi\hat{m}}-\mathbf{\hat{n}\cdot\Psi\hat{n}}
  = (\Psi_1-\Psi_3)\sin \phi^+ \sin\phi^-$,
so that from \eqref{Nmn},
\begin{equation} \label{Nsin}
\rho N^{-2}
  = \rho v^2 = (\Psi_1-\Psi_3)\sin \phi^+ \sin\phi^-,
\end{equation}
where $\phi^\pm$ are the angles that the normal $\mathbf{a}$ to the
plane of $\mathbf{S}$ makes with the optic axes.

We summarize the situation.

To determine all $\mathbf{S}^+$ and $\mathbf{S}^-$ corresponding to
\eqref{CpsiC}, the ``optics axes'' $\mathbf{h^\pm}$ are first
determined.
Then the plane of $\mathbf{C} = \mathbf{\hat{m}} + i\mathbf{\hat{n}}$
is chosen; it has unit normal $\mathbf{a}$.
If $\mathbf{a}$ is not coplanar with the two optic axes, then the
corresponding $\mathbf{\hat{m}}$ and $\mathbf{\hat{n}}$ are in the
directions of $\mathbf{r^\pm}$ given by \eqref{mn1} and $N$  is
given by \eqref{Nmn}.
If $\mathbf{a}$ is coplanar with the optic axes $\mathbf{h^\pm}$ but
not along either of them, then the corresponding
$\mathbf{\hat{m}}$ and $\mathbf{\hat{n}}$ are given by \eqref{mn2}
and $N$  is given by \eqref{Nsin}.
If $\mathbf{a}$ is along $\mathbf{h^+}$ or $\mathbf{h^-}$,
there is no propagating wave.

\subsection{Example: Crystal with a plane of symmetry}

To illustrate the method described above, we take a
``special'' material with a
symmetry plane at $x_3=0$, say.
Then the tensor $\mathbf{\Psi}_L$ defined in \eqref{Psi_L}
for the longitudinal wave is given by
\begin{equation}
  \mathbf{\Psi}_L =
     \begin{bmatrix}
     \halft d_{11} &  d_{16}         &  0 \\
     d_{16}          & \halft d_{22} &  0 \\
     0                  & 0                   &  \halft d_{33}
    \end{bmatrix}.
\end{equation}
Its eigenvalues are
\begin{equation}
\Psi_{1,3} =
   \quart \left[d_{11}+d_{22}
           \pm \sqrt{(d_{11}-d_{22})^2 + 16d_{16}^2}\right],
\quad
  \Psi_2 = \halft d_{33}.
\end{equation}
Here we assume that the stiffnesses of the material
are such that these eigenvalues are ordered
$\Psi_1 > \Psi_2 > \Psi_3$.
The corresponding unit eigenvectors are
\begin{equation} \label{e1e2e3}
  \mathbf{e}_1 =
    \frac{1}{\delta} \begin{bmatrix}
                             \halft d_{22} - \Psi_1 \\
                              -d_{16} \\
                                 0
                            \end{bmatrix},
\quad
  \mathbf{e}_2 =
    \begin{bmatrix}
      0 \\
      0 \\
      1
   \end{bmatrix},
\quad
  \mathbf{e}_3 =
    \frac{1}{\delta} \begin{bmatrix}
                              -d_{16} \\
                             \halft d_{11} - \Psi_3 \\
                                 0
                            \end{bmatrix},
\end{equation}
where $\delta$ is the positive quantity given by
\begin{equation}
\delta^2 = \textstyle{\frac{1}{8}}(d_{11}-d_{22})^2
    + 2d_{16}^2
      + \quart(d_{11}-d_{22})\sqrt{(d_{11}-d_{22})^2 + 16d_{16}^2}.
\end{equation}
With this choice, the optic axes defined by \eqref{optic} lie
in the symmetry plane $x_3=0$.

For simplicity, we now focus on waves polarized in the
symmetry plane, that is, we choose the normal $\mathbf{a}$
to the plane of $\mathbf{S}$ to be along $\mathbf{e}_2$.
Then $\mathbf{a}$ is obviously not coplanar with the optic
axes (Case (i) of the previous subsection).
Specifically, the angles between $\mathbf{a}$ and the
optic axes are $\phi^+ = \phi^- = \pi/2$.
It follows from \eqref{mn1} that $\mathbf{\hat{m}}$ and
$\mathbf{\hat{n}}$ are in the directions of
$\mathbf{h^+} \pm \mathbf{h^-}$, i.e.
\begin{equation} \label{e1e3}
\mathbf{\hat{m}} = \mathbf{e}_1, \quad
\mathbf{\hat{n}}= \mathbf{e}_3.
\end{equation}

We conclude that the following CPLIPW may propagate in a 
monoclinic crystal with symmetry plane at $x_3=0$ 
and with stiffnesses satisfying \eqref{cond2},
\begin{equation}
\mathbf{u} =
   \text{e}^{- k \mathbf{e}_3 \mathbf{\cdot x}}
    \{\mathbf{e}_1 \cos k (\mathbf{e}_1 \mathbf{\cdot x} - vt)
         - \mathbf{e}_3 \sin k (\mathbf{e}_1 \mathbf{\cdot x} - vt)\},
\end{equation}
where the orthogonal unit vectors $\mathbf{e}_1$, $\mathbf{e_3}$
are defined in \eqref{e1e2e3},
$k$ is an arbitrary real wave number,
and the real speed $v$ is given by
\begin{equation}
\rho v^2 = \Psi_1 - \Psi_3
   = \halft \sqrt{(d_{11} - d_{22})^2 + 16 d_{16}^2}.
\end{equation}

\section{Crystals with symmetries}

In this section we investigate how the conditions \eqref{cond2}
for an anisotropic crystal to admit a CPLIPW for all
choices of the polarization plane are affected when the crystal
presents certain symmetries.

\subsection{Monoclinic crystals}

Here we consider crystals with a plane of symmetry, at $x_3=0$ say.
For that class of materials,
\begin{equation} \label{mono}
d_{14} = d_{15} = d_{24} = d_{25} = d_{34} = d_{35}
  = d_{46} = d_{56} = 0.
\end{equation}
It follows that four out of the nine equations \eqref{cond2} reduce
to trivial identities, automatically satisfied.
The nine conditions  \eqref{cond2} reduce to a set of 
\textit{five} equations,
\begin{align}
& d_{16} = d_{26} = d_{36} + 2d_{45},
\nonumber \\
& 4d_{44} = d_{22} + d_{33} - 2d_{23}, \;
    4d_{55} = d_{33} + d_{11} - 2d_{13}, \;
    4d_{66} = d_{11} + d_{22} - 2d_{12}.
\end{align}

\subsection{Orthorhombic crystals}

Orthorhombic crystals possess three symmetry planes,
at $x_1=0$, $x_2=0$, $x_3=0$.
In addition to \eqref{mono},  the relations
\begin{equation} \label{ortho}
d_{16} = d_{26} = d_{36} = d_{45} = 0,
\end{equation}
also hold.
The set of nine conditions  \eqref{cond2} now reduces to a set of
\textit{three} equations,
\begin{equation}
   4d_{44} = d_{22} + d_{33} - 2d_{23}, \;
    4d_{55} = d_{33} + d_{11} - 2d_{13}, \;
    4d_{66} = d_{11} + d_{22} - 2d_{12}.
\end{equation}

\subsection{Trigonal, tetragonal, and cubic crystals}

For \textit{trigonal crystals}, $d_{24} = -d_{14} = -d_{56} \ne 0$,
and one of the conditions \eqref{cond2}, namely:
$d_{24} = d_{14} + 2d_{56}$, cannot be satisfied.

For \textit{tetragonal crystals},
$d_{11} = d_{22}$, $d_{13} = d_{23}$, $d_{44} = d_{55}$, and
the condition: $4d_{66} = d_{11} + d_{22} - 2d_{12}$
reduces to: $d_{66} = (d_{11} - d_{12})/2$,
which would mean that the crystal is in fact hexagonal (transversally
isotropic).

For \textit{cubic crystals},
$d_{11} = d_{22} = d_{33}$, $d_{12} = d_{23} = d_{13}$,
$d_{44} = d_{55} = d_{66}$, and
the condition: $4d_{66} = d_{11} + d_{22} - 2d_{12}$
reduces to: $d_{66} = (d_{11} - d_{12})/2$,
which would mean that the material is in fact isotropic.

We conclude that there are no trigonal, no tetragonal,
and no cubic crystals in which CPLIPWs 
may propagate for all orientations of the slowness plane.

\subsection{Hexagonal crystals}

For  hexagonal crystals, the following relations hold for the
stiffnesses,
\begin{equation}
d_{11} = d_{22}, \quad
d_{13} = d_{23}, \quad
d_{44} = d_{55}, \quad
d_{66} = (d_{11} - d_{12})/2,
\end{equation}
in addition to \eqref{mono} and \eqref{ortho}.
The nine equations \eqref{cond2} reduce to a \textit{single} equation,
\begin{equation} \label{hexa}
d_{44} = \textstyle{\frac{1}{4}}(d_{11} + d_{33} - 2 d_{13}).
\end{equation}

\subsection{Isotropic materials}

For isotropic materials, the following relations hold for the
stiffnesses,
\begin{equation}
d_{11} = d_{22} = d_{33}, \quad
d_{12} = d_{23} = d_{13}, \quad
d_{44} = d_{55} = d_{66} = (d_{11} - d_{12})/2,
\end{equation}
in addition to \eqref{mono} and \eqref{ortho}.
Then, the nine equations \eqref{cond2} are all identically satisfied.
However, the propagation condition \eqref{N_L}, giving the
complex scalar slowness $N_L$ now simplifies to
\begin{equation}
\rho N_L^{-2}(\mathbf{C}) = \halft d_{11}(C_1^2 + C_2^2 + C_3^2)
   = 0,
\end{equation}
for  isotropic slownesses.
It follows that CPLIPWs may not propagate in an isotropic
material, for any choice of isotropic slowness.

On the other hand, this analysis shows that ``longitudinal''
\textit{static} exponential solutions with an isotropic slowness
bivector always exist for linear isotropic elastic materials.
This result may be checked directly in the following manner.
Recall the classical equations
of equilibrium of an isotropic material,
\begin{equation}
(\lambda + \mu) u_{j,ij} + \mu u_{i,jj} = 0,
\end{equation}
where $\lambda$ and $\mu$ are the Lam\'e constants.
Then it is easy to check that the field
\begin{equation}
u_i = S_i e^{i \omega S_j x_j},
\quad
S_k S_k =0,
\end{equation}
is indeed an exact solution.

We sum up the situation.
There are no isotropic, cubic, trigonal, or tetragonal elastic 
crystals such that CPLIPWs may propagate in every plane. 
However, the propagation of CPLIPWs is 
theoretically possible for all planes of some 
triclinic, monoclinic, or orthorhombic elastic crystals 
provided some relations 
among the elastic stiffnesses are satisfied.
We note that there is unfortunately insufficient data for such
crystals available at present to enable us present an explicit 
example, but we recall that the values of the elastic stiffnesses 
change with pressure, temperature, prestress, etc. and that they may 
consequently be adjusted to produce an adequate crystal.




\begin{thebibliography}{99}


\bibitem{Borg55}
  F.E. Borgnis, 
  Specific directions of longitudinal wave propagation in anisotropic
  media.
  \emph{Physical Review}
  \textbf{98} (1955) 1000--1005.

\bibitem{Hada03}
 J. Hadamard,
 \textit{Le\c{c}ons sur la propagation des ondes et
 les \'{e}quations de l'hydrodynamique},
 Hermann, Paris (1903).

\bibitem{BoHa93}
   Boulanger, Ph. \& Hayes, M.
   \emph{Bivectors and Waves in Mechanics and Optics}
   (Chapman \& Hall, London, 1993).

\bibitem{Haye84}
   Hayes, M.
   Inhomogeneous plane waves,
   \emph{Archive for Rational Mechanics and Analysis}
   \textbf{85} (1984) 41--79.

\bibitem{DeHa02}
   Destrade, M. \& Hayes, M.
   Circularly-polarized plane waves in a deformed Hadamard material,
   \textit{Wave Motion}
   \textbf{35} (2002) 289--309.

\bibitem{DeHa04}
   Destrade, M. \& Hayes, M.
   Inhomogeneous ``longitudinal'' plane waves
   in a deformed elastic material,
   \textit{Journal of Elasticity}
   \textbf{75} (2004) 147--165.

\bibitem{Voig10}
   Voigt, W.
   \textit{Lehrbuch der Kristallphysik}
   (B.G.Teubner, Leipzig, 1910).

\bibitem{Ting00}
   Ting, T.C.T.
   Anisotropic elastic constants that are structurally invariant,
   \textit{Quarterly Journal of Mechanics and Applied Mathematics}
   \textbf{53} (2000) 511--523.

\end{thebibliography}
\end{document}